\documentclass{book}
\usepackage[ngerman,german,english]{babel}
\usepackage{makeidx}     
                         \makeindex 
\usepackage{epsfig}      
\usepackage{graphicx}    
\usepackage{Generic}     
\usepackage{amssymb, amsmath} 
\usepackage{array}       
\usepackage[sort&compress]{natbib}
\addtocounter{secnumdepth}{2} 
\makeatletter 
  \renewcommand{\paragraph}{\@startsection
      {paragraph}
      {4}
      {0.1pt}
      {0.1pt}
      {0.1pt}
      {\normalsize\rm}}
  \makeatother
\begin{document}

\renewcommand{\bibname}{References} 
\pagenumbering{arabic} 

\chapter{Elementary processes in gas discharges}
\label{BF:Elementary_Processes}

Franz~X. Bronold and Holger Fehske, {\small  Institut f\"ur Physik, 
         Ernst-Moritz-Arndt-Universit\"at Greifswald, 
         Felix-Hausdorff-Str. 6, D--17489 Greifswald, Germany\\
         (\texttt{bronold(fehske)@physik.uni-greifswald.de})}

\vspace{1cm}
\noindent
 
\section{Introduction}
\label{BF:section_0}
The plasma chemistry of reactive gas discharges 
depends strongly on the collision (elementary) processes operating, for given 
external control parameters, between the constituents of the discharge:
electrons, ions, molecules, and atoms. Mathematically, 
these processes are encoded in the collision integrals of the Boltzmann-Poisson 
equations (see Chapter XY), which are classical equations describing the discharge
on the macroscopic length scales defined by the mean-free paths of the various 
species and the screening length. The dynamics of 
the elementary processes, however, takes place on the much shorter 
microscopic scale. It is thus controlled by quantum mechanical principles which 
will be discussed in this chapter.

\begin{figure}
\centering
\includegraphics[width=.6\textwidth]{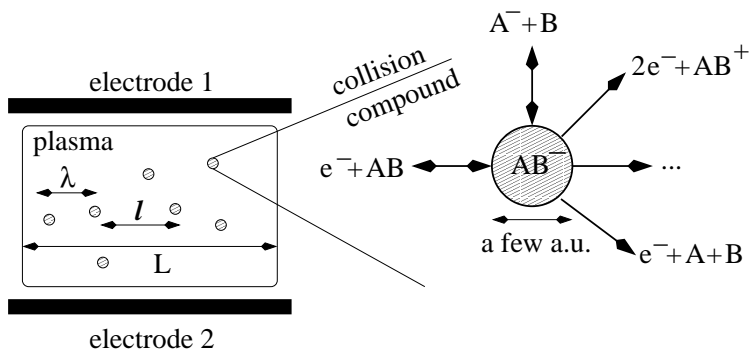}
\caption{Schematic representation of the length scales in the bulk of a typical 
low-temperature gas discharge ($n_e=n_i=10^{10}~cm^{-3}$, $T_e=10~eV$, $T_i=0.03~eV$,
and $n_{\rm gas}=10^{16}~cm^{-3}$). The macroscopic scales on which the Boltzmann-Poisson 
system operates are the spatial extension of the plasma $L\sim {\cal O}(cm)$, the 
mean-free-path of the species $l\sim {\cal O}(10^{-2}~cm)$, and the plasma screening 
length $\lambda\sim {\cal O}(10^{-3}~cm)$. On the rhs is an illustration of how the
(temporary) collision compound $AB^-$, whose length scale is a few atomic 
units, that is, of the order of $10^{-7}cm$, controls the first six reactions 
of Table~\ref{Bronold:tab:Reactions}. The branching of the $AB^-$ state at a 
given energy determines the cross sections for the reactions. The double arrow 
indicates that the fragments can appear in the entrance and the exit channel
of a collision.
}
\label{Bronold:fig:Scales}
\end{figure}
From the quantum-mechanical point of view, see Fig.~\ref{Bronold:fig:Scales} 
for an illustration, collisions proceed through temporary compound  
states~\cite{Fano81} which form and fragment on the atomic length and 
time scales~\footnote{If not stated otherwise, in this chapter it is implicitly 
assumed that physical quantities are measured in atomic units, that is,
lengths, masses, and charges are given in terms of the Bohr radius $a_B$, 
the electron mass $m_e$, and the elementary charge $e$, respectively.
The unit of energy is then $2R_0$, where $R_0$ is the Rydberg energy.},
in contrast to the macroscopic scales on which typical plasma phenomena 
such as sheath formation or wave propagation occur.
That the scales required to complete the momentum, energy, and
particle~\footnote{The term \textquotedblleft particle
\textquotedblright~denotes
either a single \textquotedblleft fundamental\textquotedblright~particle, 
having no internal degrees of freedom relevant for the collision process, 
or a collection of fundamental particles in a bound state; in the latter 
event the term \textquotedblleft fragment\textquotedblright~will be also
used.}
transfer (rearrangement) defining a collision are well separated from the 
plasma scales has two profound implications: (i) cross sections measured, 
for instance, in crossed beam or in swarm experiments, can be used without 
modifications as input data for plasma modeling and (ii) theoretical 
calculations of cross sections can neglect the plasma environment, which is
a tremendous simplification. 

Whereas the physics of a collision is unaffected by the plasma, the reverse
is of course not true. Elementary processes determine to a large extend the 
properties of a plasma (see previous chapters). In the first place, a 
gas discharge can be only maintained electrically, because electron impact 
ionization and, for an electro-negative gas, dissociative electron attachment 
produce positively and negatively charged carriers. Equally important are 
electron impact excitation and dissociation of molecules. They transfer not 
only external electric energy into internal energy but, most importantly, 
they also produce the species which are eventually utilized in the 
technological application of the discharge: electronically excited species
when the discharge is used as a light source or laser and reactive fragments
(radicals) when the discharge is employed for surface processing or catalysis.
\begin{table}
\caption{Typical collisions and their compound states for an electro-negative,
diatomic gas. The indices $i$ and $f$ denote initial and final internal
states of the molecules, atoms, and ions before and after the collision,
respectively.
}
\begin{center}
\begin{tabular}{ll}
collisions with compound state $AB^-$                    & \\
(1) ${\rm e^- + AB_i \rightarrow AB^- \rightarrow e^- + AB_f}$ & elastic and inelastic scattering\\
(2) ${\rm e^- + AB_i \rightarrow AB^- \rightarrow 2e^- + AB_f^+}$ & electron impact ionization\\
(3) ${\rm e^- + AB_i \rightarrow AB^- \rightarrow A_f^- + B_f}$ & dissociative attachment\\
(4) ${\rm e^- + AB_i \rightarrow AB^- \rightarrow e^- + A_f + B_f}$ & dissociation\\
(5) ${\rm A^- + B_i \rightarrow  AB^- \rightarrow e^- + A_f + B_f}$ & direct detachment\\
(6) ${\rm A^- + B_i \rightarrow  AB^- \rightarrow e^- + AB_f}$ & associative detachment\\
collisions with different compound states                        &  \\
(7) ${\rm e^- + A_i^- \rightarrow  A^{2-} \rightarrow 2e^- + A_f}$ & impact detachment\\
(8) ${\rm A_i^- + AB_i^+ \rightarrow  A_2B \rightarrow A_f + AB_f}$ & ion-ion annihilation\\
(9) ${\rm A_i^- + AB_i^+ \rightarrow  A_2B \rightarrow A_f + A_f + B_f}$ & ion-ion annihilation\\
(10) ${\rm A_i^- + AB_i \rightarrow  A_2B^- \rightarrow e^- + A_f + AB_f}$ & direct detachment\\
(11) ${\rm A_i^- + AB_i \rightarrow  A_2B^- \rightarrow e^- + A_2B_f}$ & associative detachment\\
(12) ${\rm e^- + AB_i^+ \rightarrow  AB \rightarrow A_f + B_f}$ & dissociative recombination\\
\end{tabular}
\end{center}
\label{Bronold:tab:Reactions}
\end{table}

Technologically interesting gas discharges contain complex molecular gases
such as $CF_4$, $CF_3I$, $C_3F_8$, $CCl_2F_2$, or $SF_6$ with
a multitude of excited and fragmented species. Even simple diatomic molecules,
for instance, $O_2$ or $N_2$ give rise to a large number of species, with an 
accordingly large number of elementary processes. Leaving aside elementary
processes containing three particles in the entrance channel, which are 
only relevant at rather high densities, the most 
common collisions for a generic electro-negative, diatomic gas are 
shown in Table~\ref{Bronold:tab:Reactions} where they are also classified 
according to the collision compounds controlling the microphysics: $AB^-$, 
the compound illustrated on the rhs of Fig.~\ref{Bronold:fig:Scales}, 
$A^{2-}$, $A_2B$, $A_2B^-$, and $AB$. The branching of the compounds
and thus the probabilities for the various collisions (cross sections), 
depend on the initial energy and on the properties of the compound states 
at the distance where they have to lock-in into the asymptotic scattering
states defining the various collisions. 

In view of the great technological and economical impact of reactive gas 
discharges, it is somewhat surprising, that the number of experimental groups 
measuring cross sections for gases of plasma-chemical relevance is rapidly 
diminishing. Thus, any listing of currently available cross section data 
must be necessarily incomplete. Since, in addition, the gases of interest
change with time, cross section data are not included at all in this chapter. 
As far as they exist, they can be found, for instance, in the review
article by Brunger and Buckman~\cite{BB02}, the monograph by Christophorou 
and Olthoff~\cite{Christophorou04}, and in web-based cross section compilations
sponsored by national research institutions and university groups whose activities
depend on atomic and molecular collision data. The largest ones, maintained,
respectively, by the International Atomic Energy Agency, the American
National Institute of Standards and Technology, by Oak Ridge National
Laboratory, USA, the Japanese National Institute for Fusion Science,
the Weizmann Institute of Science in Israel, and the Universit\'{e}
Paris-Sud in Orsay, France, are~\cite{Morgan00}:

1. http://www.iaea.org/programmes/amdis

2. http://physics.nist.gov/PhysRefData/contents.html

3. http://www-cfadc.phy.ornl.gov

4. http://dbshino.nifs.ac.jp

5. http://plasma-gate.weizmann.ac.il/

6. http://gaphyor.lpgp.u-psud.fr

Due to lack of empirical cross section data, theoretical calculations 
become increasingly important. There are two major lines of attack. One, 
the ab-initio approach, either attempts to construct parameter-free eigenstates
for collision compounds and to match these states to the asymptotic scattering 
states representing specific collision products~\cite{BT05}, or to directly 
obtain parameter-free scattering matrices from a suitable variational 
principle~\cite{WM00a}. Both approaches are extremely time consuming and,
so far, have been applied only to a few selected collisions involving 
molecules with at most three atoms. Even when the formal problems preventing 
the application to complex molecules are solved in the future, it is hard 
to imagine a non-expert routinely using an ab-initio package to generate 
cross sections for as yet unstudied collision processes.

The other, perhaps more valuable approach, as far as plasma-chemical 
applications are concerned, is the semi-empirical approach. Examples 
of which are the binary-encounter~\cite{Bethe30,Vriens69,KR94} and the
Deutsch-M\"ark~\cite{DBM00,DSB04,DSM05} model for electron impact ionization,
the resonance model for dissociative attachment~\cite{OMalley66,BHM66},
electron-detachment~\cite{Bardsley67,Herzenberg67},
dissociative recombination~\cite{Bardsley68a,Bardsley68b},
and vibrational excitation~\cite{BHM66,DH79}, and 
the Landau-Zener model for ion-ion annihilation~\cite{Olson72,OSB71}. 
Common to all these models is that they attempt to encapsulate 
the complicated collision dynamics in a few parameters which can 
be either obtained from electronic structure calculations or from 
experiments. To some extend these models are ad-hoc, but they 
have the great virtue to provide clear physical pictures for the 
collision processes, thereby helping the plasma physicist to develop 
an intuitive understanding of the microphysics of the discharge.

Despite the great diversity of elementary processes in a gas discharge, 
this chapter tries to give an uniform presentation of the subject within
the framework of multi-channel scattering theory~\cite{Gerjuoy58,Newton66,Sitenko91}
focusing, in particular, on inelastic and reactive collisions. On purpose,
cross section formulae are not developed to a point where they could be 
applied to particular collision processes, because this involves
mathematically rather technical (approximate) solutions of the given 
equations which moreover have to be worked out case-by-case. The details
can be found in the original literature. Additional information,
in particular with regards to formal aspects of calculating cross sections, 
can be extracted from the review articles by Rudge~\cite{Rudge68},
Bardsley and Mandl~\cite{BM68}, Inokuti~\cite{Inokuti71}, Lane~\cite{Lane80},
Delos~\cite{Delos81}, Chutjian and coworkers~\cite{CGW96}, Hahn~\cite{Hahn97}, 
and Florescu-Mitchell and Mitchell~\cite{FMM06}.

\section{Fundamental concepts}
\label{BF:section_1}

\subsection{Collision cross section}
\label{BF:subsection_11}

The kinetic description of a gas discharge is based on a set of Boltzmann equations 
for the distribution functions $f_i({\bf r},{\bf v},t)$ of the participating species
$i=1,2,...,N$ and those parts of Maxwell's equations which are necessary to describe
the electro-magnetic driving of the discharge. For a capacitively coupled 
radio-frequency discharge, for instance, the equations are
\begin{eqnarray}
\big[\partial_t+{\bf v}\cdot\nabla_r -
\frac{q_i}{m_i}\nabla_r\Phi\cdot\nabla_v\big]f_i({\bf r},{\bf v},t)&=&\sum_pI^p_i[\{f_j\}]~,
\label{BP1}\\
\Delta_r\Phi&=&-\frac{1}{\epsilon_0}\sum_{i=1}^N q_i \int d{\bf v} f_i~,
\label{BP2}
\end{eqnarray}
where $I^p_i[\{f_j\}]$ is the collision integral due to process $p$ appearing 
in the Boltzmann equation for species $i$ and depending on the subset $\{f_j\}$ 
of the distribution functions; $\Phi$ is the electric potential~\footnote{In 
this subsection atomic units are not used.}.
The external driving is in this case simply encoded in the boundary conditions
for the Poisson equation; $q_i$ and $m_i$ are the charge and mass of species 
$i$, respectively, and $\epsilon_0$ is the dielectric constant of the plasma.

The total number $N$ of species required for the kinetic modeling of a generic
electro-negative, diatomic gas discharge, can be deduced from 
Table~\ref{Bronold:tab:Reactions}. Clearly, besides electrons and molecules of 
the feed-stock gas $AB$, atoms $A$ and $B$, and ions $AB^+$ and $A^-$ have to be 
taken into account. In general, it is also necessary to include for 
the molecular and atomic constituents some excited (meta-stable) states. Thus, 
in total, $N=N_{AB}+N_{A}+N_{B}+N_{AB^+}+N_{A^-}+1$ species have to considered, 
where $N_\sigma$ is the number of excited states kept for the atomic or 
molecular constituent $\sigma$. 

Each reaction shown in Table~\ref{Bronold:tab:Reactions} gives rise to collision
integrals in the Boltzmann equations of the respective reaction educts (species 
in the entrance channel) and reaction products (species in the exit channel). For 
elastic scattering, the structure of the collision integral is similar to the one
originally derived by Boltzmann~\cite{B1872}. The majority of collisions, 
however, is inelastic and reactive, some involve even more than two reaction 
products. The corresponding collision integrals have then to be constructed from 
scratch, using elementary statistical considerations. 

\begin{figure}
\centering
\includegraphics[width=.4\textwidth]{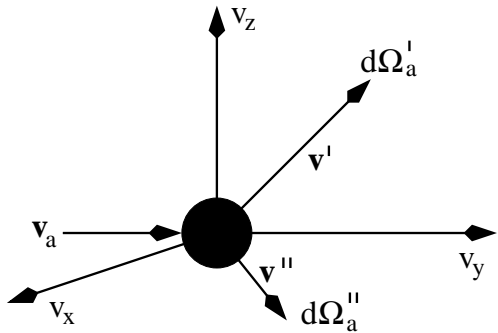}
\caption{Velocity space kinematics of electron impact ionization.
Upon impact of the incident electron with velocity vector ${\bf v}_a$ two
electrons are ejected with velocity vectors ${\bf v}'$ and ${\bf v}''$, respectively.
The solid angles $d\Omega_a'$ and $d\Omega_a''$ corresponding to these two velocities
are defined with respect to the direction of ${\bf v}_a$. Because of the large mass
difference of electrons and molecules, the molecule can be assumed to be at rest.
}
\label{Bronold:fig:Ioniz}
\end{figure}

Take, for instance, electron impact ionization, where a primary electron scatters off
a molecule and produces a secondary electron and a positively charged ion. Assuming 
the target molecule to be at rest, which is a very good approximation
because the molecule is much heavier then the electron, the collision integral
in the electron Boltzmann equation due to this process reads~\cite{WW79}
\begin{eqnarray}
I^I_e[f_e]&=&c_{AB}\bigg\{\int d{\bf v}_ad{\bf v}'' W^I({\bf v}_a;{\bf v},{\bf v}'')f_e({\bf v}_a)
+ \int d{\bf v}_ad{\bf v}' W^I({\bf v}_a;{\bf v}',{\bf v})f_e({\bf v}_a)
\nonumber\\
&-&\int d{\bf v}'d{\bf v}'' W^I({\bf v};{\bf v}',{\bf v}'')f_e({\bf v})\bigg\}~,
\end{eqnarray}
where momentum conservation is taken into account, the variables 
${\bf r}$ and $t$ in the distribution functions are suppressed, and $c_{AB}$ is the 
concentration of the molecules. 
The function $W^I({\bf v}_a;{\bf v}',{\bf v}'')$ is the probability for an incident 
electron with velocity ${\bf v}_a$ to produce two departing electrons with velocities 
${\bf v}'$ and ${\bf v}''$, respectively. Using energy conservation, it can be related
to the differential cross section for electron impact ionization
\begin{eqnarray}
W^I({\bf v}_a;{\bf v}',{\bf v}'')d{\bf v}'d{\bf v}''&=&\frac{1}{2}v_a
\sqrt{\frac{v_a^2-v''^2-2E_i/m_e}{v'^2+v''^2}}
q^{I}(v_a,v'',\Omega_a',\Omega_a'')\nonumber\\
&\times&\delta(\sqrt{v'^2+v''^2}-\bar{u})
d\Omega_a'd\Omega_a''dv'dv''~,
\end{eqnarray}
with $E_i$ the ionization energy of the molecule, $m_e$ the electron mass, 
$\bar{u}=\sqrt{v_a^2-2E_i/m_e}$, and $q^{I}(v_a,v'',\Omega_a',\Omega_a'')$ 
the differential ionization cross section defined by the relation 
\begin{eqnarray}
d\sigma^{I}=q^{I}(v_a,v'',\Omega_a',\Omega_a'')d\Omega_a'd\Omega_a''dv''
\end{eqnarray}
with $0\le v''\le\sqrt{v_a^2-2E_i/m_e}$ (cf. Fig.~\ref{Bronold:fig:Ioniz} for 
notational details).

The sub-$nm$ physics of impact ionization, that is, the momentum, energy
and electron transfer during this particular electron-molecule collision, 
is concealed in the function $q^I(v_a,v'',\Omega_a',\Omega_a'')$. 
This statement holds for all processes. The information required about 
elementary processes reduces thus to a set of (differential) cross sections. 
The direct measurement of which is tedious, expensive, and, when meta-stable 
states are involved, which cannot be prepared outside the plasma, sometimes 
even impossible. It is at this point, where theoretical calculations of cross
sections -- ab-initio or otherwise -- have a great impact.

A systematic solution of the electron Boltzmann equation expands the electron
distribution function in terms of velocity space spherical harmonics.
The angles $\Omega_a'$ and $\Omega_a''$ can then be integrated out leading to 
an hierarchy of Boltzmann equations for the expansion coefficients~\cite{WW79}.
When the anisotropy of the electron distribution is negligible, the lowest 
order equation suffices. The collision integral of interest is then
\begin{eqnarray}
q^{I}(v_a,v'')=\int d\Omega_a'd\Omega_a''q^{I}(v_a,v'',\Omega_a',\Omega_a'')~.
\end{eqnarray}
One more integration would lead to the total ionization cross section,
\begin{eqnarray}
q^{I}(v_a)=\int dv'' q^{I}(v_a,v'')~,
\end{eqnarray}
which is usually sufficient for a particle-based simulation of the ionization 
process (see Chapter XY and Ref.~\cite{MST07}). 

\subsection{Formal scattering theory}
\label{BF:subsection_12}

Collisions affecting the charge balance and chemistry of gas discharges change
the type, and sometimes even the number, of scattering fragments. An example
is electron impact ionization, the process used in the previous section to 
introduce the concept of a collision cross section. It is an inelastic, break-up 
collision with an electron and a molecule in the entrance and two electrons and 
a positive ion in the exit channel. In comparison to the entrance channel, the 
kinetic energy of the relative motion of the fragments in the exit channel is 
moreover reduced by the ionization energy of the molecule. Thus, a microscopic 
description of electron-impact ionization, and likewise of many of the other 
processes listed in Table~\ref{Bronold:tab:Reactions}, cannot be based on simple 
potential scattering theory (elastic scattering). A generalized scattering theory 
is rather required, capable to account for changes in the internal energy 
(inelasticity), for rearrangement, and for break-up of the scattering fragments.
\begin{figure}
\centering
\includegraphics[width=.5\textwidth]{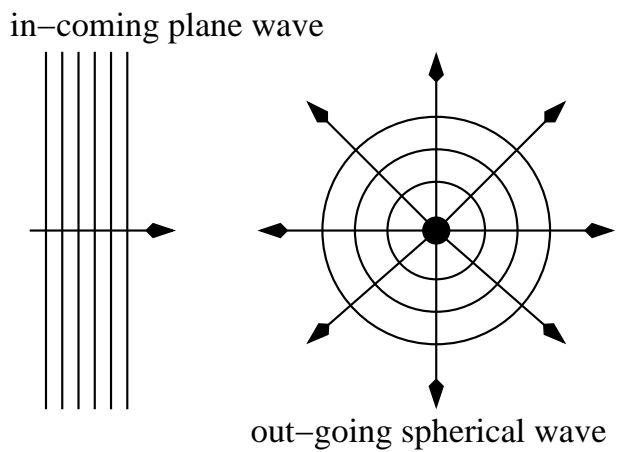}
\caption{Schematic representation of the incoming plane wave and the diverging 
spherical wave as described by the Lippmann-Schwinger equation~(\ref{LSE}).
}
\label{Bronold:fig:ScattBC}
\end{figure}

The appropriate theoretical framework is quantum-mechanical multichannel 
scattering theory~\cite{Gerjuoy58,Newton66,Sitenko91}. To introduce its essential 
ingredients, two colliding fragments are considered. In the 
center-of-mass frame, the total Hamiltonian of the system is 
\begin{eqnarray}
H=T_{\rm rel}+H_{\rm int}+V=H_0+V~, 
\label{Hentrance}
\end{eqnarray}
where $T_{\rm rel}$ is the kinetic energy of the relative motion, $H_{\rm int}$ 
controls the internal degrees of freedom of both fragments, and $V$ is the interaction 
energy between the two. The Lippmann-Schwinger equation for the scattering state with 
the boundary conditions shown in Fig.~\ref{Bronold:fig:ScattBC} reads in Dirac's bra-ket 
notation
\begin{eqnarray}
|\Psi_{{\bf k}\alpha}^{(+)}\rangle=
|\Phi_{{\bf k}\alpha}\rangle + \frac{1}{E-H_0+i\eta}V|\Psi_{{\bf k}\alpha}^{(+)}\rangle~,
\label{LSE}
\end{eqnarray}
where the first term denotes the incoming plane wave in the entrance channel. The 
channel state, $|\Phi_{{\bf k}\alpha}\rangle=|\phi_\alpha\rangle|{\bf k}\rangle$, 
satisfies $(E-H_0)|\Phi_{{\bf k}\alpha}\rangle$. Thus, 
$E=E_{{\bf k}\alpha}=k^2/(2\mu)+\omega_\alpha$, where ${\bf k}$ is the 
relative momentum, $\omega_\alpha$ is the internal energy, and $\mu$ is the reduced
mass of the fragments in the entrance channel. 

Quite generally, the scattering amplitude, which in turn determines the differential 
collision cross section, is defined as the amplitude of the out-going spherical wave 
emerging from the rhs of Eq.~(\ref{LSE}) for large inter-particle distances. Hence, 
in order to find the scattering amplitude, Eq.~(\ref{LSE}) has to be expressed 
in coordinate representation, which is here specified by ${\bf r}$, the inter-particle 
distance, and $\rho$, the internal coordinates of both particles, and then the limit 
$r\rightarrow\infty$ has to be taken. Normalizing continuum states on the momentum
scale leads then to~\cite{Newton66,Sitenko91}
\begin{eqnarray}
\Psi_{{\bf k}\alpha}^{(+)}({\bf r},\rho)&\sim& 
\sum_\beta\bigg[\exp[i{\bf k}\cdot{\bf r}]\delta_{\alpha\beta} 
-f({\bf k}'\beta,{\bf k}\alpha)\frac{\exp[ik'r]}{r}\bigg]\phi_\beta(\rho)\\
&=&\sum_\beta\bigg[\exp[i{\bf k}\cdot{\bf r}]\delta_{\alpha\beta}
+\tilde{\psi}^{(+)}_{{\bf k}'\beta,{\bf k}\alpha}(r)\bigg]\phi_\beta(\rho)
~~~\mbox{for}~r \rightarrow \infty~,
\label{asymp}
\end{eqnarray}
where ${\bf k}'$ is the relative momentum after the collision, and the prefactor in 
front of the diverging spherical wave,
\begin{eqnarray}
f({\bf k}'\beta,{\bf k}\alpha)=
-\frac{\mu}{2\pi}\langle\Phi_{{\bf k}'\beta}|V|\Psi^{(+)}_{{\bf k}\alpha}\rangle~,
\label{fscatt}
\end{eqnarray}
is the scattering amplitude. 

The differential collision cross section is microscopically defined as the ratio of
the out-going particle flux into the solid angle $d\Omega'$ originating from the 
diverging spherical wave $\tilde{\psi}^{(+)}_{{\bf k}'\beta,{\bf k}\alpha}(r)$ 
introduced in Eq.~(\ref{asymp}) and the incoming current density due to the plane 
wave. Hence,
\begin{eqnarray}
d\sigma=\frac{({\bf j}_{\rm out}\cdot{\bf e}_r) r^2d\Omega'}{k/\mu}~,
\end{eqnarray}
where ${\bf e}_r$ is the unit vector in the direction of ${\bf r}$. Inserting 
$\tilde{\psi}^{(+)}_{{\bf k}'\beta,{\bf k}\alpha}(r)$  
into the quantum-mechanical expression for the particle flux, 
${\bf j}=(1/(2i\mu))[\psi^*\nabla\psi -\psi\nabla\psi^*]$,
yields ${\bf j}_{\rm out}=k' |f({\bf k}'\beta,{\bf k}\alpha)|^2 {\bf e}_r/(\mu r^2)$
and thus
\begin{eqnarray}
d\sigma_{\alpha\rightarrow \beta}=\frac{k'}{k}
|f({\bf k}'\beta,{\bf k}\alpha)|^2d\Omega'
=\frac{\mu^2}{(2\pi)^2}\frac{k'}{k}
|\langle\Phi_{{\bf k}'\beta}|V|\Psi_{{\bf k}\alpha}^{(+)}\rangle|^2d\Omega'~,
\label{XsectM}
\end{eqnarray}
or, when continuum states are normalized on the energy scale and 
the identity $V\Psi^{(+)}=t\Phi$ is used~\cite{Newton66,Sitenko91}, 
\begin{eqnarray}
d\sigma_{\alpha\rightarrow \beta}=\frac{(2\pi)^4}{k^2}
|\langle\Phi_{{\bf k}'\beta}|t|\Phi_{{\bf k}\alpha}\rangle|^2d\Omega'~,
\label{XsectE}
\end{eqnarray}
where $t=V+V[E-H_0+i\eta]^{-1}t$ is the transition operator (T-matrix). 

In the derivation of Eq.~(\ref{fscatt}) it was implicitly assumed that ${\bf r}$ 
is the inter-particle distance in both the entrance and the exit channel. The 
particles remain therefore intact in the course of the collision. They may 
only change their internal state. Hence, the cross section formulae~(\ref{XsectM})
and~(\ref{XsectE}) can be only applied to elastic ($\omega_\beta=\omega_\alpha$) 
and inelastic $(\omega_\beta\neq\omega_\alpha$) scattering.

To obtain the cross section for reactive scattering, the fact has to be
included that the type of the particles, and hence the relative and
internal coordinates, change during the collision. Thus, writing the 
Hamiltonian in the form~(\ref{Hentrance}) is only adequate in 
the entrance channel. In the exit channel, it is more appropriate to 
partition the Hamiltonian according to the reaction products and write 
\begin{eqnarray}
H=T'_{\rm rel}+H_{\rm int}'+V'=H_0'+V', 
\label{Hexit}
\end{eqnarray}
where $T'_{\rm rel}$, $H_{\rm int}'$, and $V'$ are the relative kinetic energy, the 
internal energy, and the interaction energy in the exit channel. Then, in 
addition to Eq.~(\ref{LSE}), the scattering state
$|\Psi_{{\bf k}\alpha}^{(+)}\rangle$ 
obeys also an homogeneous Lippmann-Schwinger equation~\footnote{The 
equation is homogeneous because the incoming wave belongs to a different 
Hilbert space}, 
\begin{eqnarray}
|\Psi_{{\bf k}\alpha}^{(+)}\rangle=
\frac{1}{E-H_0'+i\eta}V'|\Psi_{{\bf k}\alpha}^{(+)}\rangle~,
\label{LSEreact}
\end{eqnarray}
from which the scattering amplitude in the exit channel (reaction amplitude)
can be deduced by the same procedure as before, except that now the 
coordinate representation with respect to ${\bf r}'$ and $\rho'$, the 
relative and internal coordinates in the exit channel, has to be chosen. 
In this representation, the scattering state becomes a diverging spherical 
wave for $r'\rightarrow\infty$. The prefactor of which  
(continuum states normalized on the momentum scale),
\begin{eqnarray}
f'({\bf k}'\beta,{\bf k}\alpha)=
-\frac{\mu'}{2\pi}
\langle\Phi'_{{\bf k}'\beta}|V'|\Psi^{(+)}_{{\bf k}\alpha}\rangle~,
\label{fexit}
\end{eqnarray} 
with $\mu'$ the reduced mass in the exit channel and 
$|\Phi'_{{\bf k}'\beta}\rangle$ an eigenstate of $H_0'$, can be identified with 
the reaction amplitude. Hence, the differential cross section for reactive
scattering is given by
\begin{eqnarray}
d\sigma_{\alpha\rightarrow \beta}=\frac{\mu}{\mu'}\frac{k'}{k}
|f'({\bf k}'\beta,{\bf k}\alpha)|^2d\Omega'
=\frac{\mu\mu'}{(2\pi)^2}\frac{k'}{k}
|\langle\Phi'_{{\bf k}'\beta}|V'|\Psi_{{\bf k}\alpha}^{(+)}\rangle|^2d\Omega'~,
\label{XsectR}
\end{eqnarray}
where energy conservation enforces now $E=k'^2/(2\mu')+\omega'_\beta=k^2/(2\mu)
+\omega_\alpha$. Obviously, Eq.~(\ref{XsectR}) reduces to Eq.~(\ref{XsectM}) for $V'=V$
which implies $H_0'=H_0$ and thus ${\bf r}'={\bf r}$, $\mu'=\mu$, and $\Phi'=\Phi$. 

If the interaction $V$ in the entrance channel is simpler than the interaction $V'$
in the exit channel, it may be more convenient to use the adjoint scattering state, 
$\langle\Psi_{{\bf k}'\beta}^{(-)}|$, which describes an incoming wave in the 
exit channel. The reaction cross section can then be written as 
\begin{eqnarray}
d\sigma_{\alpha\rightarrow \beta}=
\frac{\mu\mu'}{(2\pi)^2}\frac{k'}{k}
|\langle{\Psi}_{{\bf k}'\beta}^{(-)}|V|\Phi_{{\bf k}\alpha}\rangle|^2 d\Omega'~,
\label{XsectRadj}
\end{eqnarray}
which contains $V$ instead of $V'$.

The formalism described so far is only applicable to binary collisions, that is,
collisions containing two particles in the entrance and exit channel, respectively. 
The theoretical description of collisions involving three (or more) reaction products
(break-up collisions) can be also based on a set of Lippmann-Schwinger equations, but 
the increased dimensionality of the relative motion requires a substantial extension 
of the formalism (see Ref.~\cite{Gerjuoy58}), beyond the scope of this introductory 
presentation. Additional complications arise for 
identical particles in a given channel, or when the interaction is long-ranged, as it 
is, for instance, the case for electron-impact ionization. Ionization cross
sections are therefore hardly obtained from rigorous calculations.  
They are usually estimated from less ambitious, semi-empirical models to be 
described below.

\subsection{Adiabatic approximation}
\label{BF:subsection_13}

The formalism described in the previous subsection will be now applied to electron-molecule
scattering. To be specific, a diatomic molecule with $N$ electrons is considered.  
The scattering wavefunctions, $\Psi^{(\pm)}({\bf r}',{\bf r},{\bf R})$, from which 
the scattering amplitudes are obtained, depend then on the $N$ coordinates of the
target electrons, which are collectively denoted by ${\bf r}'$, the coordinate of
the projectile electron ${\bf r}$, and the inter-nuclear distance ${\bf R}$. In the
interest of clarity, spin degrees of freedom are suppressed.

The Lippmann-Schwinger equation for electron-molecule scattering cannot 
be solved exactly. Approximation schemes, usually based on an expansion of the 
scattering state in an appropriate basis, are therefore necessary. When the 
target states, that is, the eigenfunctions of $H^{(N)}=T_R+H_{el}^{(N)}$, where
$T_R$ is the kinetic energy of the nuclei and $H_{el}^{(N)}$ is the Hamiltonian
for the $N$ target electrons at fixed ${\bf R}$, are used, the expansion leads to 
the close-coupling approximation. Identifying $\rho$ with $({\bf r}',{\bf R})$, 
the close-coupling approximation is basically identical to 
the channel state representation described in the previous subsection. However, 
this brute-force approach is not well suited. The number of channels (electronic, 
vibrational, and rotational), which are all coupled by the Lippmann-Schwinger 
equation, is too large. In addition, due to the differences in the electronic
and nuclear energy scales, the rate of convergence is rather pure. 
\begin{figure}
\centering
\includegraphics[width=.5\textwidth]{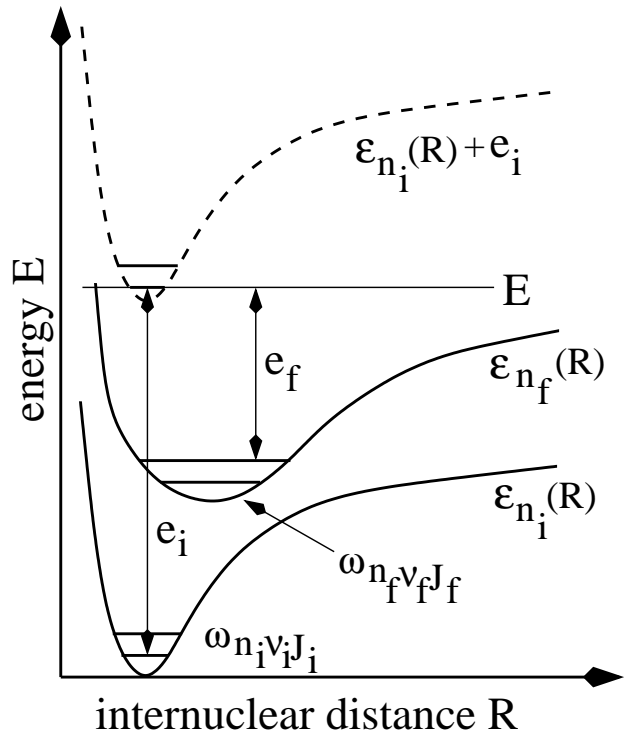}
\caption{Potential energy surface of the initial and final state of
electron-impact excitation of a diatomic molecule and the relationship of
the various energies occurring in the adiabatic approximation for the
scattering cross section. The kinetic energy of the electron before
and after the collision is denoted by $e_i=k_i^2/2$ and
$e_f=k_f^2/2$, respectively.
}
\label{Bronold:fig:Adiab}
\end{figure}

An approximation scheme accounting for the difference in the energy scales, and thus 
more appropriate for electron-molecule scattering, is the adiabatic approximation. 
It expands the scattering state in terms of eigenfunctions of $H_{el}^{(N+1)}$, 
that is, eigenfunctions of the fixed-nuclei Hamiltonian for the $N+1$ electrons
($N$ target electrons and one projectile electron) of the compound state 
$AB^-$, which depend only parametrically on the inter-nuclear 
distance. Since the dynamic coupling due to the perturbation $T_R$ is usually 
negligible, only one term contributes to the expansion.  
The computational costs are thus substantially reduced. In addition, provided
the conditions for the validity of the adiabatic approximation are fulfilled, 
the target can be also described adiabatically. Thus, the target states are 
Born-Oppenheimer states, which are of course also much easier to determine 
than the target states needed in a close-coupling calculation.

The adiabatic approximation is valid far away from excitation thresholds and when 
the collision time is much shorter than the period of nuclear vibration and 
rotation. It is thus mostly applied to non-resonant electron-molecule
scattering. As shown by Shugard and Hazi~\cite{SH75}, the differential cross 
section for electron impact excitation of a diatomic molecule reads 
for continuum states normalized on the energy scale,
\begin{eqnarray}
d\sigma_{{n_i}{\nu_i}{J_i}\rightarrow{n_f}{\nu_f}{J_f}}=\frac{(2\pi)^4}{k_i}
|T_{{n_f}{\nu_f}{J_f},{n_i}{\nu_i}{J_i}}(\Omega_f,\Omega_i)|^2 d\Omega_f~,
\label{XsectAdiab}
\end{eqnarray}
with $\Omega_i$ and $\Omega_f$ the solid angles of ${\bf k}_i$ and ${\bf k}_f$,
the electron momenta before and after the collision, respectively. The total
scattering amplitude, 
\begin{eqnarray}
T_{{n_f}{\nu_f}{J_f},{n_i}{\nu_i}{J_i}}(\Omega_f,\Omega_i)=
\int d{\bf R} F^*_{{n_f}{\nu_f}{J_f}}({\bf R})
t_{{n_f}{n_i}}(\Omega_f,\Omega_i,{\bf R})F_{{n_i}{\nu_i}{J_i}}({\bf R})~,
\label{fadiab}
\end{eqnarray}
is an average of the fixed-nuclei scattering amplitude,
\begin{eqnarray}
t_{{n_f}{n_i}}(\Omega_f,\Omega_i,{\bf R})=
\frac{k_f^{1/2}}{(2\pi)^{3/2}}\int d{\bf r}d{\bf r}'\exp[i{\bf k}_f\cdot{\bf r}]
\Phi^*_{{n_f}}({\bf r}';{\bf R})V({\bf r},{\bf r}';{\bf R})
\Psi^{(+)}_{{\cal E},\Omega,{n_i}}({\bf r},{\bf r}';{\bf R})~, 
\label{tfixed}
\end{eqnarray}
over the ro-vibrational states $F_{n\nu J}({\bf R})$ satisfying 
\begin{eqnarray}
\bigg[T_R+\epsilon_n(R)-\omega_{n\nu J}\bigg]F_{n\nu J}({\bf R})=0~,
\end{eqnarray}
where $\omega_{n\nu J}$ is the ro-vibrational energy and $\epsilon_n(R)$ is the 
potential-energy-surface defined as the eigenvalue of the fixed-nuclei
Schr\"odinger equation for the target electrons
\begin{eqnarray}
\bigg[H^{(N)}_{el}-\epsilon_n(R)\bigg]\Phi_n({\bf r}',{\bf R})=0~.
\label{HNel}
\end{eqnarray}

In view of the discussion of the previous subsection, the structure of the formulae
(\ref{XsectAdiab})-(\ref{tfixed}) should be clear. The physical meaning of the 
various energies can be deduced from Fig.~\ref{Bronold:fig:Adiab}. Note, in particular, 
that the energy ${\cal E}=\epsilon_{n_i}+k_i^2/2=\epsilon_{n_i}+E-\omega_{{n_i}{\nu_i}{J_i}}$
appearing in Eq.~(\ref{tfixed}) is only the total electronic energy, for the projectile 
electron and the $N$ target electrons, in contrast to the total energy $E$ which includes
electronic and nuclear degrees of freedom. Ro-vibrational states are characterized by the 
quantum number $n$ of the electronic state in which the nuclear motion takes place, the 
vibrational quantum number $\nu$, and the rotational quantum number $J$. Energy conservation
implies $E=k_i^2/2+\omega_{{n_i}{\nu_i}{J_i}}=k_f^2/2+\omega_{{n_f}{\nu_f}{J_f}}$. 

Within the adiabatic approximation, it is necessary to determine the fixed-nuclei 
scattering amplitude and the ro-vibrational states of the molecule, which in turn 
depend on the potential energy surface of the target molecule. The former can be 
obtained, as a function of ${\bf R}$, from the asymptotics of the $N+1$ electron, 
fixed-nuclei Lippmann-Schwinger equation, which determines 
$\Psi^{(+)}_{{\cal E},\Omega,{n_i}}({\bf r},{\bf r}';{\bf R})$, while the latter 
requires, again as a function of ${\bf R}$, the solution of the $N$ electron 
problem~(\ref{HNel}). In both cases, anti-symmetrized wavefunctions have to 
be used because of the indistinguishability of electrons. Thus, even when the 
nuclear motion is split off, the calculation of cross sections for electron-molecule
scattering remains a formidable many-body problem~\cite{Lane80}.

\subsection{Resonant scattering}
\label{BF:subsection_14}

At electron energies of a few electron volts or less the collision time is long 
compared to the period of the inter-nuclear motion and the adiabatic approximation 
fails. The projectile electron is then so slow that it is captured by the 
molecule giving rise to a bound state of the negatively charged molecular
ion, which is the collision compound for electron-molecule 
scattering~\footnote{The electronic and vibrational properties of 
negative ions play therefore an important role in low-temperature gas
discharges with $k_BT_e\sim {\cal O}(1-10eV)$, even when the gas is not electro-negative, 
that is, when the negative ion is unstable on the plasma time scale and therefore 
irrelevant at the kinetic level.}. 
This state interacts with the electron-molecule scattering continuum, acquires 
therefore a finite lifetime, and turns into a quasi-bound (auto-detaching) 
state. Auto-detaching states play a central role in (vibrational) excitation, 
attachment, recombination, and detachment collisions. From a theoretical 
point of view, all these processes can be hence analyzed within a model 
describing a discrete state (resonance) embedded in a continuum of 
scattering state.

A particularly elegant derivation of the resonance model for an electron 
colliding with a diatomic molecule containing $N$ electrons has been given by 
Domcke~\cite{Domcke91} who uses many-particle Green functions to reduce the 
$N+1$ electron scattering problem to an effective single-electron problem. 
The reduction is achieved by two projections: First, electronic states which 
are not accessible at the energy considered, are eliminated by introducing an 
optical potential for the incoming electron. Then, in a second step, the
fixed-nuclei T-matrix is split into a rapidly varying part due to the 
quasi-bound resonant state and a slowly varying background term. The splitting
of the T-matrix can be shown to be equivalent to an effective single-electron
Hamiltonian at fixed-nuclei describing a resonance coupled to a continuum
of states. At the end, the kinetic energy of the nuclei is included and the
electronic degrees of freedom are integrated out to obtain an effective 
Lippmann-Schwinger equation for the nuclear dynamics which, with appropriate 
boundary conditions, can then be used to calculate the collision cross sections
of interest.

It is essential for the formalism that the optical potential supports a resonance
and that the resonance can be extracted from the single-electron continuum such 
that the scattering background contains no spurious resonances. The many electron
problem is then completely buried in an optical potential, which can be calculated
separately using, for instance, many-body perturbation theory~\cite{Meyer89}. In
principle, the formalism can handle more than one resonance as well as electronically
inelastic processes~\cite{BCM99}. So far, however, it has been only applied to 
electronically elastic collisions involving a single resonance and a single potential 
energy surface for the nuclear motion of the target. 
\begin{figure}
\centering
\includegraphics[width=.8\textwidth]{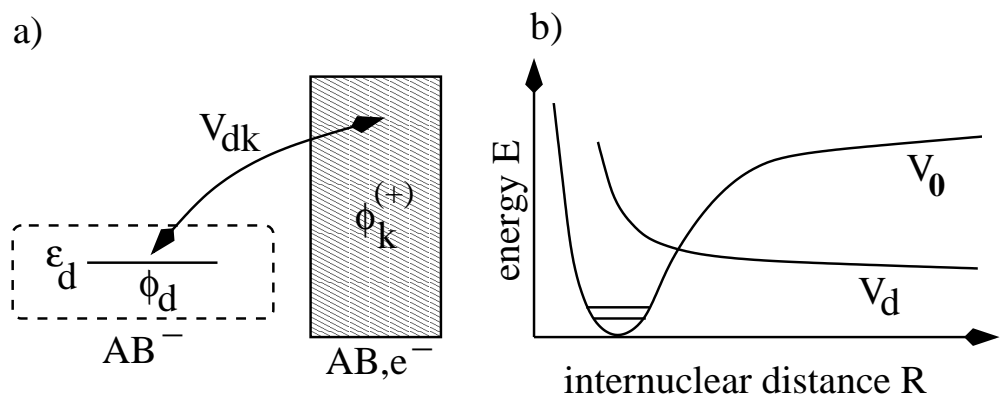}
\caption{Illustration of the effective Hamiltonian~(\ref{Heff}) defining the
resonance model.
}
\label{Bronold:fig:ResMod}
\end{figure}

The effective Hamiltonian, visualized in Fig.~\ref{Bronold:fig:ResMod}, reads 
\begin{eqnarray}
H&=&H_0+V
\label{Heff}\nonumber\\
H_0&=&|\phi_d\rangle\bigg[T_R+V_d\bigg]\langle \phi_d| + 
\int k dk d\Omega_k|\phi_{\bf k}^{(+)}\rangle 
\bigg[T_R+V_0+\frac{k^2}{2} \bigg]\langle \phi_{\bf k}^{(+)}|
\nonumber\\
V&=&\int k dk d\Omega_k\bigg[|\phi_d\rangle V_{d{\bf k}}\langle \phi_{\bf k}^{(+)}|
+ {\rm h.c.}\bigg]~,
\end{eqnarray}
where $|\phi_d\rangle$ is the resonant state and $|\phi^{(+)}_{\bf k}\rangle$ are the
scattering states orthogonal to $|\phi_d\rangle$. The coupling between the two 
types of states is given by $V_{d{\bf k}}=\langle \phi_d|H_{el}|\phi^{(+)}_{\bf k}\rangle$, 
where $H_{el}=-\nabla^2/2+V_{\rm opt}$ is the Hamiltonian for a single electron in the 
optical potential $V_{\rm opt}$; $V_d=\langle\phi_d|H_{el}|\phi_d\rangle+V_0$ is an 
operator specifying the potential of the resonant state and $V_0$ is the corresponding
operator for the potential of the target molecule. 

The Hamiltonian~(\ref{Heff}), and thus the Lippmann-Schwinger equation associated with 
it, operate in the combined Hilbert space of the scattered electron and the two nuclei. 
Projecting out the electron, switching to the coordinate representation 
with respect to the inter-nuclear distance, and using out-going wave boundary conditions 
yields, when rotations of the molecule are ignored, a Lippmann-Schwinger equation,
\begin{eqnarray}
\bigg[E+\frac{1}{2\mu}\frac{d^2}{dR^2}-V_d(R)\bigg]\Psi_{dE}^{(+)}(R)
-\int dR'F^{(+)}(R,R',E)\Psi_{dE}^{(+)}(R')=J(R)~,
\label{LSEres}
\end{eqnarray}
with a kernel
\begin{eqnarray}
F^{(+)}(R,R',E)=\int k dk d\Omega_k V_{d{\bf k}}(R) 
G^{(+)}(R,R',E-k^2/2)V^*_{d{\bf k}}(R')~,
\label{Ffct1}
\end{eqnarray}
where 
\begin{eqnarray}
G^{(+)}(R,R',E)=
\langle R|\big[E+\frac{1}{2\mu}\frac{d^2}{dR^2}-V_0(R)+i\eta\big]^{-1}|R'\rangle
\label{Gfct}
\end{eqnarray}
is the Green function for the nuclear motion on the potential energy surface of 
the target, $V_0(R)=\langle R|V_0|R\rangle$, and $\mu$ is the reduced mass of the 
nuclei; $V_d(R)=\langle R|V_d|R\rangle=\epsilon_d(R)+V_0(R)$ is the potential 
energy surface of the resonant state and 
$V_{d{\bf k}}(R)=\langle R|V_{d{\bf k}}|R\rangle$.
The inhomogeneity $J(R)$ on the rhs of Eq.~(\ref{LSEres}) depends on the
boundary conditions. It will be discussed below for particular collision 
processes. 

Equation~(\ref{LSEres}) is an effective Lippmann-Schwinger equation for the nuclear 
dynamics in the energy-dependent, nonlocal, and complex potential of the resonant state. 
To make this more explicit, $G^{(+)}_0(R,R',E)$ is expressed in terms of a complete set of 
target nuclear wavefunctions $\chi_{\nu}(R)$. Employing Dirac's identity and assuming 
that the interaction between the resonance and the scattering continuum depends only 
on $k$ (isotropic interaction), leads to 
\begin{eqnarray}
F^{(+)}(R,R';E)=\Delta(R,R';E)-\frac{i}{2}\Gamma(R,R';E) 
\label{Ffct2}
\end{eqnarray}
with 
\begin{eqnarray}
\Delta(R,R';E)=
4\pi\sum_\nu~{\rm P}\!\!\int dE'\frac{V_{dE'}(R)\chi^*_\nu(R)\chi_\nu(R')V_{dE'}^*(R')}
{E-\omega_\nu-E'}
\label{Delta}
\end{eqnarray}
and  
\begin{eqnarray}
\Gamma(R,R';E)=
8\pi\sum_\nu V_{dE-{\omega_\nu}}(R)\chi^*_\nu(R)\chi_\nu(R')V_{dE-{\omega_\nu}}^*(R')~,
\label{Gamma}
\end{eqnarray}
where the symbol "P" denotes the principal value of the integral and $\omega_\nu$ stands 
for the vibrational energies of the target molecule. 

The inverse of $\Gamma(R,R';E)$ is the lifetime of the resonance. 
As expected, the auto-detaching property of the resonance arises from its coupling
to the scattering continuum. In principle, the resonance has a finite lifetime
even in the absence of this coupling because the optical potential, $V_{\rm opt}$, 
is complex. Its imaginary part induces therefore a lifetime. This contribution,
however, is much smaller than the one due to $V_{dE}$. Hence, 
it is usually neglected. 

The non-locality of the potential complicates the numerical solution of Eq.~(\ref{LSEres}).
In the early applications of the resonance 
model~\cite{OMalley66,BHM66,Bardsley67,Herzenberg67,Bardsley68a,Bardsley68b,BM68}, 
the nonlocal potential was thus replaced by a local one. The local approximation can 
be obtained from Eqs.~(\ref{Delta})--(\ref{Gamma}) by identifying the energy available 
for the scattered electron with an effective resonance energy: $E-\omega_\nu\approx 
E_{\rm res}(R)$~\cite{Domcke91}. The completeness of the vibrational target states 
can then be used to obtain $\Delta(R,R';E)=\Delta_L(R)\delta(R-R')$ and 
$\Gamma(R,R';E)=\Gamma_L(R)\delta(R-R')$ with
\begin{eqnarray}
\Gamma_L(R)=8\pi|V_{d{E_{\rm res}(R)}}(R)|^2,~~~~~~
\Delta_L(R)=4\pi~{\rm P}\!\!\int\frac{|V_{d{E_{\rm res}(R)}}(R)|^2}{E_{\rm res}(R)-E'}~,
\end{eqnarray}
which reduces the Lippmann-Schwinger equation to an ordinary differential equation:
\begin{eqnarray}
\bigg[E+\frac{1}{2\mu}\frac{d^2}{dR^2}-V_0(R)-\Delta_L(R)+\frac{i}{2}\Gamma_L(R)\bigg]
\Psi_{dE}^{(+)}(R)=J(R)~.
\label{LSEresApp}
\end{eqnarray}

The scattering cross sections derived from Eq.~(\ref{LSEres}) or Eq.~(\ref{LSEresApp})
are only as good as the optical potential, $V_{\rm opt}$, the potential energy 
surfaces $V_0(R)$ and $V_d(R)$, and the coupling function $V_{d{\bf k}}(R)$. These
quantities have to be obtained from separate calculations, ideally using ab-initio 
techniques of molecular structure theory, or directly from experiments. But even 
then, the cross sections form the resonance model are semi-empirical in the sense 
that a priori assumptions about the relevance or irrelevance of molecular 
ion and target states have to be made from the very beginning. This weakness of 
the model, however, leads at the same time to its strength: Technically 
tractable equations, with an intuitive physical interpretation, 
which, in the local approximation, can be even solved analytically with 
semi-classical techniques~\cite{KY84,KK90a,KK90b}.

\section{Typical processes}
\label{BF:section_2}

Now representative elementary processes are discussed in more detail, focusing, 
in particular, on reactive and inelastic collisions, which change the internal 
energy and composition of the scattering fragments. Elastic scattering between 
the various species is not explicitly included although 
the associated cross sections are usually much larger than 
the cross sections for inelastic and reactive collisions combined. But they only 
randomize the directed motion of the electrons in the electric field. The thereby 
induced changes of the electron energy distribution function affects the plasma-chemistry
only indirectly, in as much as an increase of energy in the electronic subsystem makes 
certain collisions more probable than others.

\subsection{Production of ions}
\label{BF:subsection_21}

The main production processes for ions in a molecular, electro-negative gas discharge 
are electron impact ionization and dissociative electron attachment. Both processes share 
the same compound state, $AB^-$, but the former leads to positive molecular ions whereas 
the latter to negative atomic ions. Impact ionization is in addition also the main source 
for electrons which, depending on their energy, trigger a multitude of excitation and 
dissociation collisions.

{\it Electron impact ionization} is similar to electronic excitation (see below), 
except that the excited state belongs to the two-electron continuum. In the notation
of the previous subsection, the differential cross section is thus proportional to the 
modulus of the reaction amplitude squared,  
\begin{eqnarray}
d\sigma^{\rm I}_{\alpha\rightarrow \beta}
\sim |\langle\Phi'_{\beta{\bf k}_1{\bf k}_2}|V'|\Psi_{\alpha{\bf k}}^{(+)}\rangle|^2
d\Omega_1d\Omega_2dE_2~,
\end{eqnarray}
where $\Omega_1$ is the solid angle associated with ${\bf k}_1$, the momentum of 
the primary electron, $\Omega_2$ is the solid angle associated with ${\bf k}_2$,
the momentum of the secondary electron, and $E_2$ is the energy of the secondary 
electron. The exit channel state, $|\Phi'_{\beta{\bf k}_1{\bf k}_2}\rangle
=|\phi'^{\rm AB^+}_\beta \psi'_{{\bf k}_1{\bf k}_2}\rangle$, describes the internal 
state of the positive ion and the six-dimensional relative motion between the two 
electrons and the ion, which is controlled by $V'$, that is, the Coulomb 
interaction of the primary and secondary electron with each other and with the 
positive ion. The entrance channel of the set of Lippmann-Schwinger equations 
satisfied by the scattering state $|\Psi^{(+)}_{\alpha{\bf k}}\rangle$ contains 
an inhomogeneity, $|\Phi_{\alpha{\bf k}}\rangle=|\phi^{\rm AB}_\alpha\rangle |{\bf k}\rangle$,
where $|\phi^{\rm AB}_\alpha\rangle$ is the initial state of the molecule 
and $|{\bf k}\rangle$ is the plane wave representing the relative motion of the
initial electron and the molecule.

From a rigorous  mathematical point of view, the exact two-electron scattering 
state in the field of a positive ion is unknown. Hence, exact calculations of 
electron impact ionization cross sections are not available, even for atoms, 
where complications due to the nuclear dynamics are absent. Perturbative 
treatments, based, for instance, on the distorted Born approximation, which 
replaces $\psi'_{{\bf k}_1{\bf k}_2}$ by a plane wave for the primary electron 
and a Coulomb wave for the secondary electron, are possible, provided the velocity 
of the incident electron is much larger than the velocities of the electrons
bound in the molecule. Exchange and correlation effects can then be ignored and 
the calculation of cross sections for electron-impact ionization reduces essentially 
to the calculation of photo-ionization cross sections. This approach, 
due originally to Bethe~\cite{Bethe30} and reviewed by Rudge~\cite{Rudge68}
and Inokuti~\cite{Inokuti71}, fails at low electron energies. But in that region, 
Vriens' binary-encounter model~\cite{Vriens69} can be used. It assumes that the 
primary electron interacts pairwise with target electrons, leaving the remaining 
electrons and the nuclear dynamics undisturbed. The ionization cross section 
is then basically the Mott cross section for the collision of two electrons, 
appropriately modified by the binding and kinetic energies of the target 
electrons.

The semi-empirical {\it Kim-Rudd model} for electron impact ionization~\cite{KR94}
combines the Bethe model with the binary-encounter model. For 
an individual orbital, the ionization cross section is then given by 
\begin{eqnarray}
\sigma^{I}(e)=\frac{S}{e+u+1}\bigg\{\frac{1}{2}Q\bigg(1-\frac{1}{e^2}\bigg)\ln e + (2-Q)
\bigg[1-\frac{1}{e}-\frac{\ln e}{e+1}\bigg]\bigg\}~,
\label{XsectIoniz}
\end{eqnarray}
with $e=E/B$ and $u=U/B$, where $E$ is the kinetic energy of the incident electron, 
$U$ is the average kinetic energy of the orbital, and $B$ is the binding energy of
the orbital. The remaining quantities are $S=4\pi N/B^2$ and
$Q=2BM_i^2/N$ with $N$ the orbital occupation number and $M_i$ an integral 
over the differential oscillator strength defined in Ref.~\cite{KR94}. The 
results obtained from Eq.~(\ref{XsectIoniz}) are surprisingly good. 

Another widely used semi-empirical model for electron impact ionization is the 
{\it Deutsch-M\"ark model}~\cite{DBM00,DSB04,DSM05}. It uses cross sections for the 
atomic constituents 
of the molecules and sums them up according to the atomic population of the molecular 
orbitals which is obtained from a Mullikan population analysis. The total ionization 
cross section can then be written as
\begin{eqnarray}
\sigma^I(E)=\sum_{n,l,i}g_{i,nl}\pi r_{i,nl}^2 N_{i,nl} f(e_{i,nl})
\end{eqnarray}
with $r_{i,nl}^2$ the mean square radius of the $n,l$-th sub-shell of the 
constituent $i$, $N_{i,nl}$ the occupancy of that sub-shell, $g_{i, nl}$
weighting factors which have to be determined empirically using reliable 
atomic cross section data, and a function 
\begin{eqnarray}
f(e_{i,nl})=\frac{1}{e_{i,nl}}\bigg(\frac{e_{i,nl}-1}{e_{i,nl}+1}\bigg)^{3/2}
\bigg\{1+\frac{2}{3}\bigg(1-\frac{1}{2e_{i,nl}}\bigg)
\ln \big[2.7 + \sqrt{e_{i,nl}-1}\big]\bigg\}~,
\end{eqnarray}
which describes the energy dependence of the cross section. The 
quantity $e_{i,nl}=E/B_{i,nl}$ with $B_{i,nl}$ the ionization energy of the 
considered sub-shell of the $i$-th constituent. More elaborate versions of the 
Deutsch-M\"ark model account also for the different angular symmetries of the 
sub-shells~\cite{DBM00}. 

The cross sections for impact ionization are fairly large because 
it is an optically allowed process (this can be most clearly seen when the 
Born approximation is applied) and the energy in the final state can be distributed
among the two electrons in infinite many ways. Hence, a gas discharge usually
contains a high concentration of electrons and positive ions.  

Negative ions are produced by {\it dissociative attachment}, which is a resonant
process due to the collision of a slow electron with a (in general) vibrationally 
excited molecule, where the compound state $AB^-$ dissociates into a negative 
ion $A^-$ and a neutral atom $B$. In principle, it would be also conceivable 
that the incident electron is permanently bound to the molecule and a negative 
molecular ion is formed. A necessary condition for this to happen would be 
that the relevant potential energy surface of the $AB^-$ state is bonding 
and has a minimum outside the \textquotedblleft potential well 
\textquotedblright~corresponding to the electronic groundstate of the $AB$
molecule. However, the electron is always captured above 
the re-detachment threshold, that is, the molecular ion is initially in an 
excited state. It can be only stabilized when it looses or disperses excess 
energy. The former occurs through collisions with other molecules or the wall 
while the latter takes place through vibrational modes of the $AB^-$ compound. 
In most cases, these processes are not very efficient and the $AB^-$ 
state is only a short-lived resonance not affecting the macrophysics
of the discharge. 

\begin{figure}
\centering
\includegraphics[width=.5\textwidth]{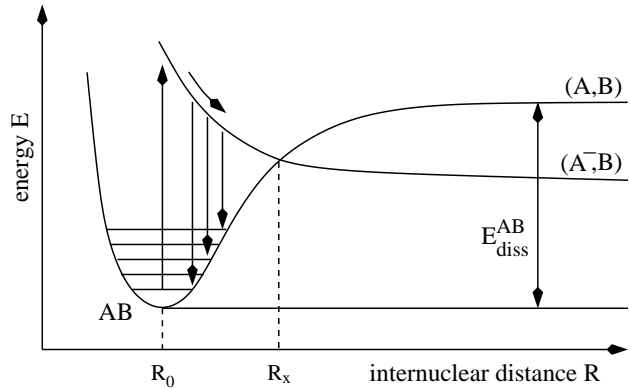}
\caption{Schematic representation of the relevant potential energy surfaces
for dissociative attachment. For simplicity, it is assumed that the molecule
$AB$ is initially in the vibrational groundstate, although this is usually
not the case. The electron is thus captured at $R=R_0$,
where $R_0$ is the equilibrium distance of the two nuclei (upward
directed vertical arrow). The thereby created $AB^-$ compound is envisaged
to be in an anti-bonding state whose potential energy surface crosses at
$R=R_x$ the potential energy surface of the $AB$ state. For $R<R_x$ the
compound state has thus a finite probability to decay (downward directed
vertical arrows). However, provided the state survives until the nuclear
distance $R>R_x$, it asymptotically reaches the $(A^-,B)$ dissociation
limit and dissociative attachment is completed.}
\label{Bronold:fig:PESforDA}
\end{figure}

As pointed out by O'Malley~\cite{OMalley66} and others~\cite{BHM66}, dissociative 
attachment most likely occurs when an anti-bonding state of the collision compound
$AB^-$ crosses the initial potential energy surface of the $AB$ molecule as 
schematically shown in Fig.~\ref{Bronold:fig:PESforDA}. This is the typical 
situation for the resonance model to be applicable. Thus, with the appropriate 
identifications and boundary conditions, the cross section for dissociative 
attachment can be calculated from the effective Hamiltonian~(\ref{Heff}). 

The natural boundary conditions for dissociative attachment are an incoming 
plane wave for the relative electron-molecule motion and an outgoing spherical 
wave in the relative $(A^-,B)$ motion. The formula for the attachment cross 
section, which is an reaction cross section [cp. with Eq.~(\ref{XsectR})],
would then however contain the interaction in the exit channel, that is, the 
interaction between $A^-$ and $B$, which is not part of the 
resonance model. This complication can be avoided when the adjoint scattering 
problem is considered, whose boundary conditions are an incoming plane wave in the 
$(A^-,B)$ channel and an outgoing spherical wave in the electron-molecule channel. 
The cross section contains then the interaction between the resonant state and the 
$e-AB$ scattering continuum. Specifying Eq.~(\ref{XsectRadj}) to the present 
case, the differential cross section for dissociative 
attachment becomes
\begin{eqnarray}
d\sigma^{\rm DA}_{\nu_i}=
\frac{mM}{(2\pi)^2}\frac{K}{k_i}
\bigg|\int dR \bigg[\Psi^{(-)}_{dE}(R)\bigg]^*V^*_{dk}(R)\chi_{\nu_i}(R)\bigg|^2 d\Omega_K~,
\label{XSforDA}
\end{eqnarray}
where $m$ and $k$ are the reduced mass and the relative momentum of the 
$(e^-,AB)$ system, respectively, $M$ and $K$ are the corresponding quantities 
in the $(A^-,B)$ system, $\chi_{\nu_i}(R)$ is the vibrational state of the 
molecule, and $\Psi^{(-)}_{dE}(R)$ is the scattering state satisfying
the complex conjugate of Eq.~(\ref{LSEres}) with $V_d(R)=V_{\rm AB^-}(R)$,
$V_0(R)=V_{\rm AB}(R)$, $V_{dk}(R)$ the interaction between the anti-bonding
$AB^-$ state and the electron-molecule scattering continuum, and 
$J(R)=[V_d(\infty)-V_d(R)]\exp[iKR]$. The total energy available for the
collision $E=k^2/2+\omega_{\nu_i}=K^2/(2M)+\omega_{A^-}+\omega_B$, 
where $\omega_{A^-}$ and $\omega_{B}$ denote the internal energies of the 
ion and atom, respectively, and $\omega_{\nu_i}$ is the vibrational energy 
of the molecule.

In order to avoid additional indices, quantum numbers for the internal 
state of the $(A^-,B)$ system are suppressed and rotations of the 
molecule are also ignored. Because the period of rotation is much longer 
than the collision time, rotations could be included within the adiabatic 
approximation. Finally, provided the kinetic energy of the incident 
electron, $k^2/2$, is much larger than the vibrational energy 
of the target, $\omega_{\nu_i}$, the local approximation could be employed, 
that is, Eq.~(\ref{LSEresApp}) could be used instead Eq.~(\ref{LSEres}). 
Further details about the calculation of electron attachment cross 
sections can be found in the review article by Chutjian and 
collaborators~\cite{CGW96}.

Bardsley and coworkers~\cite{BHM66} have shown that within the semi-classical 
approximation $d\sigma_{\nu_i}/d\Omega_K$ factorizes into a capture cross 
section, which describes the formation of the compound state, and a survival 
probability for that state. The semi-classical calculation moreover shows that 
the capture cross section increases faster with temperature than the survival
probability decreases. Thus, the attachment cross section increases with 
temperature. It is however always one or two orders of magnitude smaller 
than the cross section for the corresponding elastic electron-molecule 
scattering. 

\subsection{Destruction of ions}
\label{BF:subsection_22}

Whereas in an electro-negative gas discharge only two processes are primarily
responsible for the production of ions and electrons -- electron impact 
ionization and dissociative attachment -- a large number of processes may 
lead to the destruction of ions and electrons (see Table~\ref{Bronold:tab:Reactions}).
Depending on the parameters of the discharge, the loss of ions may be due primarily
to recombination or detachment. Recombination may furthermore occur between positive
ions and electrons or between positive and negative ions (annihilation). For 
molecular positive ions, the former process is usually accompanied by dissociation
and is thus called dissociative recombination. Ion-ion annihilation, on the other 
hand, is in most cases only a charge transfer, which does not lead to 
a rearrangement of the nuclei. Finally, detachment of negative ions can be either 
initiated by electrons or by neutrals. Electron-induced detachment is similar to 
electron impact ionization whereas detachment due to (molecular) neutrals 
may lead to dissociation, as well as, when the neutral is in a meta-stable 
state, to association. 
\begin{figure}
\centering
\includegraphics[width=.5\textwidth]{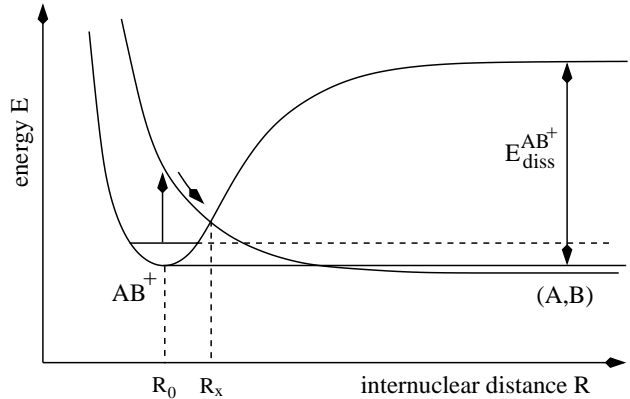}
\caption{Dissociative recombination occurs when an anti-bonding potential
energy surface of the (excited) molecule, $V_{AB}(R)$, crosses the potential 
energy surface of the positive molecular ion $AB^+$. Although the states 
involved are different, the mechanism is similar to dissociative attachment. 
When the molecule, which is the collision compound in this case, 
survives auto-detachment for $R<R_x$, it may reach the $(A,B)$ 
dissociation limit. As a result, dissociative recombination takes place.
}
\label{Bronold:fig:PESforDR}
\end{figure}

{\it Dissociative recombination} is triggered by slow electrons. It is thus a resonant process, 
similar to dissociative attachment. Very often, it is the dominant loss process for 
positive ions and electrons with relatively large cross sections because at least one 
of the atomic fragments in the exit channel is usually in an excited state
implying that the energy in the exit channel can be distributed in many different 
ways~\cite{FMM06}. 

The resonance model for dissociative recombination, originally proposed 
by Bardsley~\cite{Bardsley68a,Bardsley68b}, 
is based on the potential energy surface diagram shown in Fig.~\ref{Bronold:fig:PESforDR}. 
The most favorable situation for the process to take place is when an anti-bonding 
potential energy surface of the molecule $AB$, which is the collision compound for 
this reaction, crosses the potential energy surface of the positive ion and is for 
large inter-nuclear distances below the vibrational ground state of the ion. This is, 
for instance, the case for $H_2^+$, $N_2^+$, and $O_2^+$.  

The cross section for dissociative recombination can then be casted into 
the form~(\ref{XSforDA}), with $M$ and $K$ the reduced mass and the relative 
momentum of the $(A,B)$ system, respectively, $m$ and $k$ the corresponding 
quantities of the $(e^-,AB^+)$ system, $\chi_{\nu_i}(R)$
the initial vibrational state of $AB^+$, and $\Psi^{(-)}_{dE}(R)$ the 
solution of the complex conjugate to Eq.~(\ref{LSEres}) with $V_d(R)=V_{AB}(R)$, 
$V_0(R)=V_{AB^+}(R)$, and $V_{dk}(R)$ the interaction between the $AB$ state and
the $e^--AB^+$ scattering continuum. The inhomogeneity representing the boundary 
condition, $J(R)=[V_d(\infty)-V_d(R)]\Phi^C_{K}(R)$, contains now a Coulomb wave, 
$\Phi^C_K(R)$, because the scattering continuum is for two charged particles: 
an electron and a positive ion. More sophisticated approaches use quantum-defects
to characterize the scattering states~\cite{Giusti80,AKO82}. 

\begin{figure}
\centering
\includegraphics[width=.5\textwidth]{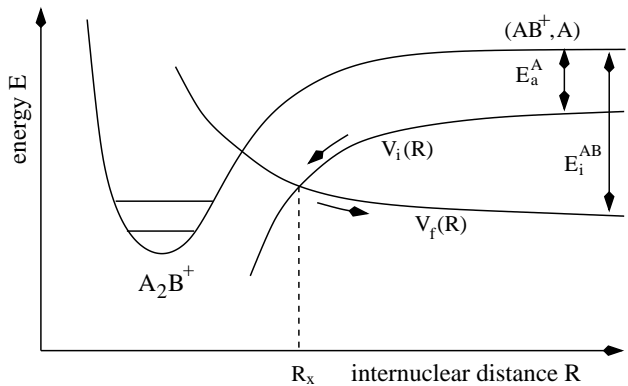}
\caption{Relation of the potential energy surfaces typical for ion-ion annihilation. 
In the entrance channel, the nuclear dynamics is governed by $V_i(R)$, the potential 
energy surface of the $(AB^+,A^-)$ configuration, whereas the dynamics in the exit 
channel is controlled by $V_f(R)$, which is the potential energy surface of the 
$(AB,A)$ configuration. The 
strong mixing of the two configurations encoded in $V(R)$ (see Eq.~(\ref{lambda})) 
is maximal at $R=R_x$. Both, $V_i(R)$ and $V_f(R)$, belong to the collision 
compound $A_2B$. The potential energy surface for the $A_2B^+$ molecule is 
also shown. It allows to define the relevant energies: the ionization energy of 
the $AB$ molecule, $E_{ion}^{AB}$, and electron affinity of the $A$ atom, $E_{a}^A$.
}
\label{Bronold:fig:PESforAnni}
\end{figure}

In electro-negative gas discharges, there is an additional recombination channel:
{\it Ion-ion annihilation}. The potential energy surface diagram relevant for this process
is shown in Fig.~\ref{Bronold:fig:PESforAnni}. The essential point is that at an 
inter-nuclear spacing $R=R_x$ the potential energy surface for the $(A^-,AB^+)$ 
configuration, which decreases with decreasing $R$ because of the Coulomb attraction
between the two ions, crosses an anti-bonding potential energy surface of the $A_2B$ 
collision compound. At this separation, the probability for the system to switch from
the $(A^-,AB^+)$ to the $(A,AB)$ configuration is particularly high. Based on this 
observation, a semi-empirical Landau-Zener model can be constructed which relates the 
cross section for ion-ion annihilation to the ionization energy of the $AB$ molecule and 
the electron affinity of the $A$ atom (see Fig.~\ref{Bronold:fig:PESforAnni}).

The {\it Landau-Zener model} illustrates quite nicely how semi-empirical models 
encode complicated processes in a few physically intuitive parameters. 
In contrast to the processes discussed so far, where 
the relevant nuclear dynamics took place on a single potential energy surface,
ion-ion annihilation forces the nuclei to switch 
between two potential energy surfaces of the collision compound. The minimal 
theoretical model is therefore based on two coupled Lippmann-Schwinger equations
for the relative motion of the nuclei. The Landau-Zener model is the semi-classical 
approximation to this set of equations. Following Olson~\cite{Olson72}, the 
total ion-ion annihilation cross section is then given by 
\begin{eqnarray}
\sigma^{\rm IIA}(E)=4\pi R_x^2\bigg[1+\frac{\Delta E}{E}\bigg]F(\lambda)~,
\label{XSforAnni}
\end{eqnarray}
with 
\begin{eqnarray}
F(\lambda)=\int_1^\infty dx~x^3\exp(-\lambda x)\bigg[1-\exp(-\lambda x)\bigg]
\end{eqnarray}
and
\begin{eqnarray}
\lambda=2\pi\sqrt{\frac{M}{2}}\frac{|V(R_x)|^2}{|V_i'-V_f'|\sqrt{E+\Delta E}}~,
\label{lambda}
\end{eqnarray}
where $E$ and $M$ are the kinetic energy and the reduced mass of the relative motion
of the $(A^-, AB^+)$ system, $\Delta E$ is the energy gain due to annihilation, $V(R_x)$ 
is the interaction between the $(A^-, AB^+)$ and $(A, AB)$ configurations at $R=R_x$, 
and $V_{i,f}'=dV_{i,f}(R_x)/dR$ with $V_i(R)$ and $V_f(R)$ the potential energy 
surfaces of the $(A^-,AB^+)$ and $(A,AB)$ system, respectively. 

Equation~(\ref{XSforAnni}) can be developed further, by recalling that $V_i(R)\sim R^{-1}$ 
(Coulomb interaction between $A^-$ and $AB^+$) and $V_f(R)\sim r^{-n}$ with 
$n>1$ (polarization interaction between $A$ and $AB$). Hence, 
for large enough $R_x$, $|V_i'-V_f'|\approx R_x^{-1}$ and $\Delta E\approx 
-U_i(R_x)=R_x^{-1}$. Usually, $\Delta E\gg E$. Combining all this leads to 
$\lambda\approx \sqrt{2M}\pi R_x^{5/2} |V(R_x)|^2$~\cite{Olson72}. Since 
$F(\lambda)$ approaches its maximal value $F_{max}\approx 0.1$ at 
$\lambda_{max}\approx 0.424$, a rough estimate for the annihilation cross 
section is
\begin{eqnarray}
\sigma^{\rm IIA}(E)\approx 1.3 R_x^2\bigg[1+\frac{1}{R_xE}\bigg]
\end{eqnarray}
with $R_x$ determined from $\lambda_{max}=\sqrt{2M}\pi R_x^{5/2} |V(R_x)|^2$ or 
from empirical cross section data for high energies where $E\gg R_x^{-1}$ and 
$\sigma(E)\rightarrow 1.3R_x^2$. To determine $V(R_x)$ is not trivial. Ideally, 
it can be parameterized in terms of an effective ionization energy $E^{AB}_{i}$ of 
the $AB$ molecule and the electron affinity $E^A_a$ of the $A$ atom. 
Olson and coworker~\cite{OSB71} obtain for large inter-nuclear distances
$V(R)=1.044\sqrt{E_i^{AB}E_a^A}\sqrt{q}\bar{R}\exp[-0.857\bar{R}]$ with 
$\bar{R}=0.5(\sqrt{2E^A_a}+\sqrt{2E^{AB}_i})R$ and $q$ the Franck-Condon factor 
that represents the overlap of the vibrational states (see below). In order to 
apply this formula, the ionization potential and the electron affinity have to be 
known. For atomic negative ions this is no problem. But for molecular negative 
ions this can be subtle because the ion may be in an (unknown) excited state. 
Fortunately, molecular negative ions have a rather short lifetime, as discussed
before. They are thus not of major concern in the commonly used gas discharges.

For molecular positive ions, {\it dissociative ion-ion annihilation} may be also 
possible. It contains two relative motions in the exit channel and is thus 
much harder to analyze theoretically. Experimentally, it is not always 
possible to discriminate between this process and the charge-transfer-type reaction 
discussed in the previous paragraphs. The energy dependence of the two cross 
sections is however essentially the same, except that they reach the 
constant value at different energies. Thus, when $R_x$ is determined from
the high energy asymptotics of empirical data, the annihilation cross 
section is in fact an effective cross section, comprising both 
charge-transfer-type and dissociative ion-ion annihilation.

\begin{figure}
\centering
\includegraphics[width=.5\textwidth]{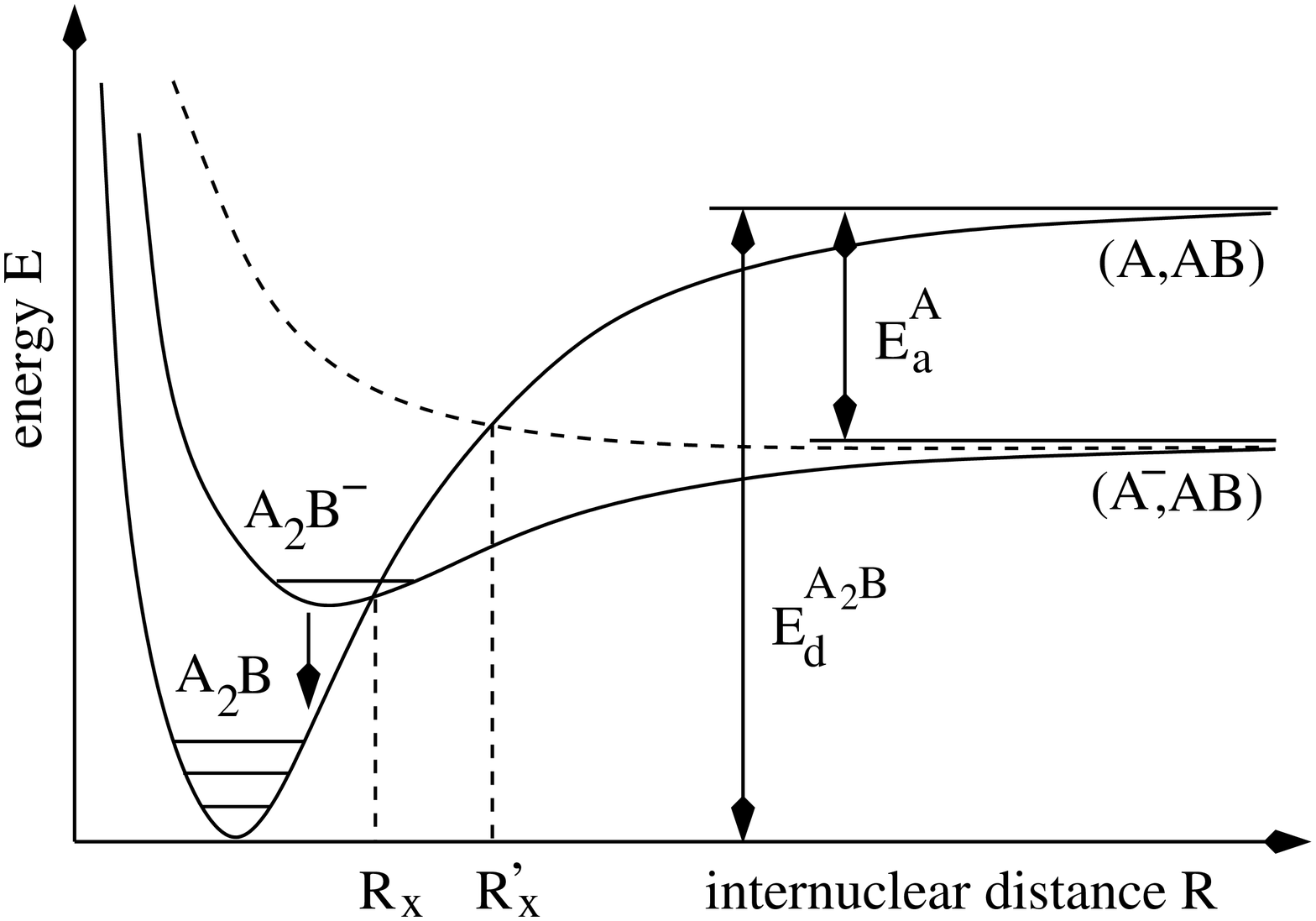}
\caption{Illustration of associative detachment. The potential
energy surface of a bonding state of the collision compound
$A_2B^-$ crosses the potential energy surface of the $A_2B$ molecule.
Auto-detachment of the compound state produces then an electron and a
vibrationally excited $A_2B$ molecule. Also shown is an anti-bonding
$A_2B^-$ state (dashed line). Detachment may also occur from such a state,
but then the initial kinetic energy of the $(A^-,AB)$ system has to be
larger than $V_{A_2B^-}(R_x')-V_{A_2B^-}(\infty)$.
}
\label{Bronold:fig:PESforAD}
\end{figure}

In contrast to positive ions, negative ions can be also destroyed by slow 
collisions with neutral particles (detachment). To be specific, 
{\it associative detachment} of negative ions due to molecules is discussed. 
Microscopically, it is a resonant process similar to dissociative 
attachment and dissociative recombination. An auto-detaching, resonant 
state plays thus an important role. In Fig.~\ref{Bronold:fig:PESforAD} 
the relevant potential energy surfaces are shown. The asymptotically 
stable configuration $(A^-,AB)$ changes with decreasing 
$R$ into a quasi-bound, resonant collision compound, $A_2B^-$, from
which the electron auto-detaches, leaving behind a free electron 
and a neutral particle in an excited vibrational 
state~\cite{Herzenberg67,Bardsley68a,Bardsley68b}. 

The differential cross section for this process can be obtained from 
the resonance model, using as boundary conditions an incoming plane
wave in the $(A^-,AB)$ channel and an outgoing spherical wave in 
the $(e^-,A_2B)$ channel. Normalizing continuum states on the energy 
scale and applying Eq.~(\ref{XsectR}) leads to 
\begin{eqnarray}
d\sigma^{\rm AD}_{\nu_f}=
\frac{(2\pi)^4}{K^2}
\bigg|\int dR \chi^*_{\nu_f}(R)V^*_{dk}(R)\Psi^{(+)}_{dE}(R)\bigg|^2 d\Omega_{k_f}~,
\label{XSforAD}
\end{eqnarray}
where $\chi_{\nu_f}(R)$ is the vibrational state of the $A_2B$ molecule
and $\Psi^{(+)}_{dE}$ satisfies Eq.~(\ref{LSEres}) with $V_d(R)=V_{A_2B^-}(R)$, 
$V_0(R)=V_{A_2B}(R)$, and $V_{dk}(R)$ the interaction between the quasi-bound 
$A_2B^-$ state and the $e^--A_2B$ scattering continuum. The inhomogeneity 
is $J(R)=[V_d(\infty)-V_d(R)]\Phi_K(R)$ with $\Phi_K(R)$ a plane wave for the 
relative motion of the $(A^-,AB)$ system. Energy conservation  
enforces $E=K^2/(2M)+V_d(\infty)=k^2/(2m)+\omega_{\nu_f}$ where
$M$ and $m$ are the reduced masses of the $(A^-,AB)$ and the 
$(e^-,A_2B)$ system, respectively. Notice, since 
$d\sigma^{\rm AD}_{\nu_f}$ is a differential cross section for reactive 
scattering, the interaction in the exit channel, $V_{dk}$, appears 
in Eq.~(\ref{XSforAD}), as it should be.

The most favorable situation for associative detachment is when the potential
energy surface of the neutral molecule in the exit channel supports a bound 
state whose dissociation energy $E_{\rm d}^{A_2B}$ is larger than the 
electron affinity $E_{\rm a}^A$ of the $A$ atom. This situation is shown 
in Fig.~\ref{Bronold:fig:PESforAD}. Provided detachment is mediated by an attractive
state of the collision compound, it takes place even for vanishing initial 
kinetic energy in the $(A^-, AB)$ channel. If this is the case, detachment
is a very efficient loss channel for negative ions, even at low temperatures.
If, on the other hand, the compound state is anti-bonding, associative detachment
occurs only when the initial kinetic energy of the colliding particles is 
larger than $V_d(R_x')-V_d(\infty)$, where $R_x'$ is the point where the 
repulsive potential energy surface crosses $V_0(R)$. 

\subsection{Excitation of internal degrees}
\label{BF:subsection_23}

The chemistry and charge balance of a gas discharge is also affected by 
inelastic collisions, that is, collisions which increase the internal energy of 
molecules, atoms, and ions. Excited (meta-stable) particles are reactive and 
participate in basically all particle number changing collisions. In low 
temperature gas discharges, vibrationally excited molecules play a particularly 
important role, because, for typical operating conditions, they are efficiently 
produced by resonant electron-molecule scattering. 

The kinetic energy of electrons in a low-temperature gas discharge is typically
a few electron volts. At these energies, the electron-molecule collision 
time is rather long, favoring therefore resonant enhancement of the collision. 
The cross section for {\it vibrational excitation} of molecules is thus given by 
Eq.~(\ref{XsectM}) with the scattering state obtained from the resonance model. 
Specifically for continuum states normalized on the energy scale, the 
cross section becomes 
\begin{eqnarray}
d\sigma^{\rm VE}_{\nu_i\rightarrow\nu_f}=
\frac{(2\pi)^4}{k_i^2}
\bigg|\int dR \chi_{\nu_f}^*(R)V_{d{\bf k}_f}^*(R)\Psi^{(+)}_{dE}(R)\bigg|^2
d\Omega_f~,
\end{eqnarray}
where $\chi_{\nu_f}(R)$ is the vibrational state of the molecule after the 
collision and $\Psi^{(+)}_{dE}(R)$ is the solution of the effective Lippmann-Schwinger 
equation~(\ref{LSEres}) with $V_d(R)=V_{AB^-}(R)$, $V_0(R)=V_{AB}(R)$, 
$V_{dk}(R)$ the interaction between the resonant $AB^-$ state and the 
electron-molecule scattering continuum, and $J(R)=V_{d{\bf k}_i}(R)\chi_{\nu_i}(R)$.
Energy conservation implies $E=k_i^2/(2\mu)+\omega_{\nu_i}=
k_f^2/(2\mu)+\omega_{\nu_f}$ with $\mu$ the reduced electron-molecule mass.
\begin{figure}
\centering
\includegraphics[width=.6\textwidth]{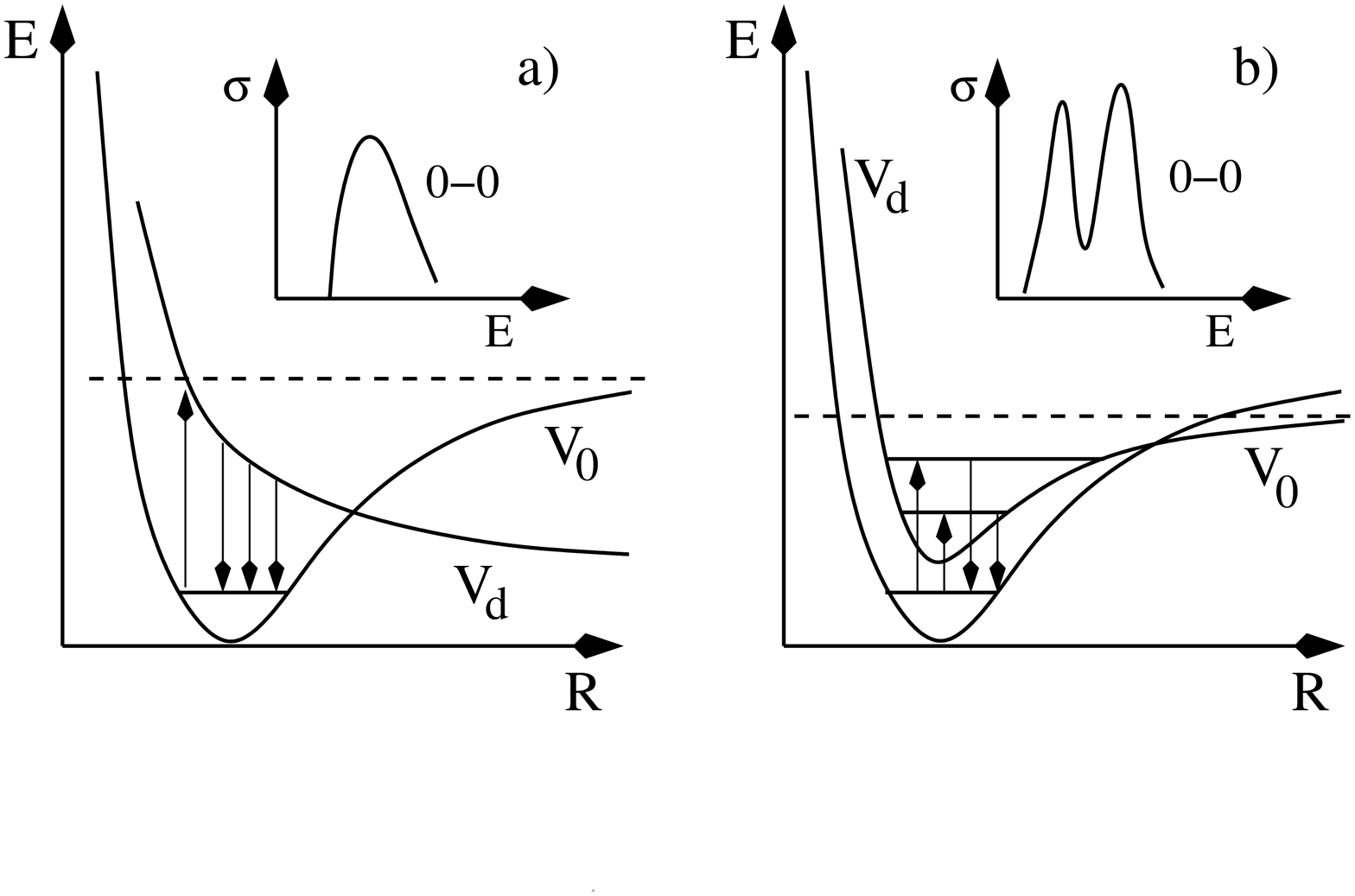}
\caption{Illustration of resonant vibrational excitation of a molecule for
a large (a) and a small (b) imaginary part of $F(R,R';E)$. The dashed line
indicates the dissociation threshold, $V_0(R)$ is the potential energy surface for 
the molecule, $V_d(R)$ is the potential energy surface for the auto-detaching resonant
state of the negatively charged molecule, and the insets show the typical shape of 
the cross section for the two cases.
}
\label{Bronold:fig:VE}
\end{figure}

The shape of the vibrational excitation cross section depends on the imaginary part
of $F(R,R';E)$, that is, on $\Gamma(R,R';E)$ (see Eqs.~(\ref{Ffct1}) and~(\ref{Ffct2})). 
When it is large compared to the  inverse of the period of vibration in the resonant 
state, that is, for a short lived resonance, the cross section is smooth. For a 
long-lived resonance, that is, when $\Gamma(R,R';E)$ is much smaller than the period 
of vibration, the cross sections consist of a series of peaks reflecting the vibrational 
states of the resonant state. This dependence of the cross section on the lifetime of 
the resonance is illustrated in Fig.~\ref{Bronold:fig:VE}.

When the energy exceeds the dissociation energy, {\it dissociation} may occur, either
indirectly through the resonant state, or directly through a transition to the 
continuum of the nuclear motion of the molecule. The indirect process contributes 
only when the target is initially in a high excited vibrational state~\cite{AW93}. 
Usually, the direct process dominates. For large enough energies, the adiabatic
approximation applies and the dissociation cross section is given by 
\begin{eqnarray}
d\sigma^{\rm D}_{\nu_i\rightarrow \omega_f}=\frac{(2\pi)^4}{k_i}
|T_{{\omega_f}{\nu_i}}(\Omega_f,\Omega_i)|^2 d\Omega_f
\end{eqnarray}
with 
\begin{eqnarray}
T_{\omega_f\nu_i}(\Omega_f,\Omega_i)=
\int dR F^*_{\omega_f}(R)
t(\Omega_f,\Omega_i,R)F_{\nu_i}(R)~,
\end{eqnarray}
where $t(\Omega_f,\Omega_i,R)$ is the electronically elastic fixed-nuclei scattering 
amplitude and $F_{\omega_f}(R)$ is a continuum nuclear wavefunction of the molecule.
Electronic quantum numbers are suppressed because they are unchanged; vibrational
excitation and dissociation involve only one potential energy surface of the 
target molecule. 

Finally, the {\it production of electronically excited molecules} occurs due to 
scattering with high energy electrons. For energies far away from the dissociation 
threshold, the adiabatic approximation holds. Therefore, ignoring the rotational 
degrees of freedom, the transition amplitude can be written as 
\begin{eqnarray}
T_{{n_f}{\nu_f}{n_i}{\nu_i}}(\Omega_f,\Omega_i)&=&
\int dR F^*_{{n_f}{\nu_f}}(R)
t_{{n_f}{n_i}}(\Omega_f,\Omega_i,R)F_{{n_i}{\nu_i}}(R)
\nonumber\\
&=&t_{n_fn_i}(\Omega_f,\Omega_i,R_0)\int dR F^*_{{n_f}{\nu_f}}(R)
F_{{n_i}{\nu_i}}(R)~.
\end{eqnarray}
In the second line the fact has been utilized that $F_{{n_i}{\nu_i}}(R)$ is 
strongly peaked at $R=R_0$, where $R_0$ is the position of the minimum of 
the potential energy surface of the initial electronic state of the molecule. 
The differential cross section for electronic excitation is thus given by 
\begin{eqnarray}
d\sigma^{\rm EE}_{n_i\nu_i\rightarrow n_f\nu_f}
=\frac{d\sigma^{R_0}_{n_i\rightarrow n_f}}{d\Omega_f} 
\bigg|\int dR F^*_{n_f\nu_f}(R) F_{n_i\nu_i}(R)\bigg|^2 d\Omega_f~,
\end{eqnarray}
where the first factor is the fixed-nuclei differential cross section at 
$R=R_0$ for the electronic transition $n_i\rightarrow n_f$ and the second 
is the Franck-Condon factor $q$ already mentioned in connection with 
ion-ion annihilation.

\section{Concluding remarks}
\label{BF:section_3}

This chapter discussed elementary collision processes, as they typically occur 
in molecular gas discharges, focusing, in particular, on inelastic and 
reactive collisions which change the internal energy and composition of the 
scattering fragments: ionization, attachment, recombination, annihilation,
detachment, and excitation. These processes are the primary driving force 
of the plasma-chemistry in these discharges. Elastic scattering, on the other 
hand, is not explicitly discussed, because its main effect, depositing 
energy into the electronic subsystem, affects the plasma-chemistry only 
indirectly.

Instead of merely listing kitchen-made cross section formulae and unrelated
cross section data, emphasis has been given on an unified description of 
elementary processes based on general principles of quantum-mechanical
multi-channel scattering theory. By necessity, the presentation is rather 
dense. Technical details left out, as well as cross section data, can be found, 
respectively, in the original papers and the review articles, monographs, 
and web-sites mentioned in the introduction.

As far as the kinetic modeling of gas discharges is concerned, collision cross sections
are input parameters for the collision integrals of the Boltzmann equations. They 
are thus always convoluted with distribution functions. Some of the quantum-mechanical 
intricacies of particle en- and de-tanglement occurring during a collision 
are thus eventually averaged out. Calculating cross sections 
earmarked for plasma-chemical applications with the most sophisticated 
quantum-mechanical methods is thus not only not feasible for the (poly-atomic) 
gases of interest, it would be also overkill. Cross sections obtained from 
effective (semi-empirical) models containing only those microscopic degrees 
of freedom which, for given external control parameters, may 
eventually become active in the collision integrals, should be in fact sufficient. A 
systematic effort to develop (and solve) this type of models for the various
processes may significantly increase the predictability of plasma modeling. It 
may thus help to turn plasma processing as a whole from an art to a science~\cite{Jacob83}. 

\bibliographystyle{Processes}
\bibliography{Processes}

\printindex 

\end{document}